\documentclass[preprint]{elsarticle}
\usepackage[utf8]{inputenc}
\usepackage{amsmath,amsfonts}
\usepackage{graphicx}
\usepackage{geometry}
\usepackage{booktabs}

\journal{}

\begin{document}

\begin{frontmatter}



\title{High-order ADI scheme for option pricing in stochastic volatility models}


\author[label1]{Bertram D{\"u}ring\corref{cor1}} 
\cortext[cor1]{Corresponding author}
\address[label1]{Department of Mathematics, University of Sussex, Pevensey II, Brighton, BN1 9QH, United Kingdom.}
\ead{b.during@sussex.ac.uk}

\author[label1]{James Miles}
\ead{james.miles@sussex.ac.uk}

\begin{abstract}
\noindent We propose a new high-order alternating direction implicit (ADI) finite difference scheme 
for the solution of initial-boundary
value problems of convection-diffusion type with mixed
derivatives and non-constant coefficients, as they arise from stochastic
volatility models in option pricing.
Our approach combines different high-order spatial discretisations with Hundsdorfer and
Verwer’s ADI time-stepping method, to obtain an efficient method which
is fourth-order accurate in space and second-order accurate in time.
Numerical experiments for the European put option pricing problem using Heston’s stochastic volatility model
confirm the high-order convergence.
\end{abstract}


\end{frontmatter}



\section{Introduction}

\noindent In financial option pricing, stochastic volatility models as
the Heston model \cite{Hes93}
have become one of the standard approaches.
Unlike the classical Black \& Scholes model \cite{BlaSch73} 
the volatility (or standard deviation) of the option's underlying
asset is not assumed to be constant, but is modelled
as a second, correlated stochastic diffusion process. This additional source of
randomness allows to model option prices more accurately and to fit
higher moments of the asset return distribution. Using Ito's lemma and
standard arbitrage arguments, partial differential equations of
convection-diffusion type with mixed second-order derivatives are
derived for pricing options.

For some stochastic volatility models and under additional restrictions, closed-form
solutions can be obtained by Fourier methods (e.g.\
  \cite{Hes93,Due09}). 
Another approach is to derive approximate analytic expressions, see e.g.\
\cite{BeGoMi10} and the literature cited therein.
In general, however, ---even in the Heston model \cite{Hes93}  when the
parameters in it are non constant--- the partial differential
equations arising from stochastic volatility models have
to be solved numerically. Moreover, many (so-called American) options
feature an additional early exercise right. Then one has to solve a
free boundary problem which consists of the partial differential
equation and an early
exercise constraint for the option price. Also for this problem one
typically has to resort to numerical approximations.

In the mathematical literature, there are many papers on numerical
methods for option pricing, mostly addressing the one-dimensional case of a single
risk factor and using standard, second order finite difference
methods (see, e.g., \cite{TavRan00} and the references therein). More
recently, high-order finite difference schemes (fourth order in space)
were proposed \cite{Gupta,Rigal99,SpotzCarey} that use a compact stencil (three points in space). In
the option pricing context, see e.g.\ \cite{DuFoJu04,DuFoJu03,LiaKha09}.

There are less works considering numerical methods for option
pricing in stochastic volatility models, i.e.\ for two spatial
dimensions. Finite difference approaches that are used are
often standard, second-order methods, e.g. in \cite{IkoToi07} where different efficient
methods for solving the American 
option pricing problem for the Heston model are proposed.
Other approaches include finite element-finite
volume \cite{ZvFoVe98}, multigrid \cite{ClaPar99}, sparse wavelet
\cite{HiMaSc05}, or spectral methods \cite{ZhuKop10}.

The classical alternating direction implicit (ADI) method, introduced by
Peaceman and Rachford \cite{PeacmanRachford95}, Douglas \cite{Douglas,
Gunn}, Fairweather and Mitchell \cite{Fairweather},
is a very powerful method that is especially
useful for solving parabolic equations ({\em without\/} mixed derivative
terms) on rectangular domains. Beam
and Warming \cite{BeamWarming}, however, have shown that no simple 
ADI scheme involving only discrete solutions at time levels $n$ and $n+1$ can be
second-order accurate in time in the presence of mixed derivatives. To overcome this limitation and construct an
unconditionally stable ADI scheme of second order in time, a number of
results have been given by Hundsdorfer and Verwer \cite{Hund02, Verwer} and 
more recently by in't Hout and Welfert \cite{HouWel07}.
These schemes are second-order accurate in time and space.
In \cite{HouFou10} different second-order ADI schemes of this type are
applied to the Heston model.
In \cite{DuFoRi13} this approach was combined with different
high-order discretisations in space, using high-order compact schemes
for two-dimensional
convection-diffusion problems {\em with mixed derivatives and constant
coefficients}.
In \cite{HeEhGu15} this approach was combined with sparse grids and
applied to a multi-dimensional Black-Scholes equation, again {\em with constant
coefficients}. 

In the present paper we present a high-order ADI method for option
pricing in a rather {\em general class of stochastic volatility models},
extending the approach in \cite{DuFoRi13}. This involves two-dimensional
convection-diffusion equations {\em with mixed derivative terms and
space-dependent coefficients\/} which adds substantial algebraic
complexity in the derivation of the scheme. The new scheme is 
 second-order accurate in time and fourth-order accurate in space.

This paper is organised as follows. In the next section we discuss
stochastic volatility models for option pricing and the related
pricing partial differential equation. In Section~\ref{sec:HV} we
recall the Hundsdorfer-Verwer ADI splitting in time. For the
spatial discretisation we introduce different high-order methods, in
Section~\ref{sec:HOC} for the implicit steps, and in
Section~\ref{sec:HOexpl} for the explicit steps. The solution of the
resulting scheme and numerical boundary conditions are discussed in
Sections~\ref{sec:HOsolve} and \ref{sec:bc}.
We present numerical convergence and stability results in
Section~\ref{sec:num}.
Section~\ref{sec:conc} concludes.


\section{Stochastic volatility models}

\noindent We consider the following class of stochastic volatility models: asume that asset spot price $0 \leq S(t)< \infty$ and variance $0 \leq \sigma(t)< \infty$ follow two stochastic diffusive processes for $t\in[0,T]$,
\begin{subequations}
\label{eq:SVmodels}
\begin{align}
dS(t)&=\bar{\mu} S(t)dt + \sqrt{\sigma (t)} S(t) dW^{(1)}(t), \\
d\sigma(t)&=\widetilde{\kappa}(\sigma(t))^\alpha(\widetilde{\theta}-\sigma(t))dt + v(\sigma(t))^\beta dW^{(2)}(t) ,
\end{align}
\end{subequations}
which are characterised by two Brownian motions, $dW^{(1)}(t)$ and
$dW^{(2)}(t)$, with constant correlation parameter
$dW^{(1)}(t)dW^{(2)}(t)=\rho dt$. 
The drift coefficient for stochastic asset returns is given by the mean return of the asset where $\bar{\mu}\in \mathbb{R}$ and the diffusion coefficient is given by $\sqrt{\sigma (t)}S(t)$. 

The drift coefficient of the asset variance is given by
$\widetilde{\kappa}(\sigma(t))^\alpha(\widetilde{\theta}-\sigma(t))$,
where constants $\widetilde{\kappa} \geq 0$ and $\widetilde{\theta}
\geq 0$ are the mean reversion speed of $\sigma(t)$ and the long run
mean of $\sigma(t)$, respectively. The diffusion
coefficient is given by $v(\sigma(t))^\beta$ where constant $v\geq 0$
is the volatility of volatility. The
constant riskless interest rate is denoted by $r \geq 0.$ 
The constants $\alpha,\beta$ 
determine the stochastic volatility model used. 

The class of stochastic volatility models \eqref{eq:SVmodels} includes
a number of known stochastic volatility models:
The most prominent stochastic volatility model, the \textit{Heston
  model} \cite{Hes93} (also called \textit{square root (SQR) model}) specifies the variance by
$$d\sigma(t) = \widetilde{\kappa} \left(\widetilde{\theta} -\sigma(t) \right) {\rm d}t + v \sqrt{\sigma(t)} {\rm d}W^{(2)}(t).$$
Other known stochastic volatility models include the \textit{GARCH}
(or \textit{VAR model}) model, see \cite{Duan95}, where the stochastic variance is modelled by
$$d\sigma(t) = \widetilde{\kappa} \left(\widetilde{\theta} -\sigma(t) \right) {\rm d}t + v \sigma(t) {\rm d}W^{(2)}(t),$$
and the \textit{3/2 model} \cite{Lewis00} in which the variance follows the process
$$d\sigma(t) = \widetilde{\kappa} \left(\widetilde{\theta} -\sigma(t) \right) {\rm d}t + v \sigma^{\frac{3}{2}}(t) {\rm d}W^{(2)}(t).$$
All of the three  stochastic volatility models mentioned above use a linear mean-reverting drift for the stochastic process of the variance $v(t)$, 
but there are also models, in which the drift is mean reverting in a
non-linear fashion.
Following \cite{ChJaMi08}, we denote these models with an additional ``N'':
in the \textit{SQRN model} the stochastic variance follows
$$dv = \widetilde{\kappa} \sigma(t)\left(\widetilde{\theta} -\sigma(t) \right) {\rm d}t + v \sqrt{\sigma(t)} {\rm d}W^{(2)}(t) ,$$
in the \textit{VARN model}
$$dv = \widetilde{\kappa} \sigma(t)\left(\widetilde{\theta} -\sigma(t) \right) {\rm d}t + v \sigma(t) {\rm d}W^{(2)}(t) ,$$
and in  the \textit{$3/2$-N model}
$$dv = \widetilde{\kappa} \sigma(t)\left(\widetilde{\theta} -\sigma(t) \right) {\rm d}t + v \sigma^{\frac{3}{2}}(t) {\rm d}W^{(2)}(t) ,$$
see \cite{ChJaMi08}.

Applying standards arbitrage arguments and Ito’s lemma to the
class of stochastic volatility models \eqref{eq:SVmodels}, we can
derive the following second order partial differential equation for
any financial derivative $V(S,\sigma,t)$, to be solved backwards in
time with $0 < S < \infty $, $0 < \sigma < \infty$, $t \in [0,T)$:
\begin{equation}
\label{PDE}
V_{t} + \frac{S^2\sigma}{2}V_{SS} + \rho v \sigma^{\beta+\frac{1}{2}} S V_{S\sigma} + \frac{v^2\sigma^{2\beta}}{2}V_{\sigma\sigma} + rSV_{s} +[\widetilde{\kappa}\sigma^{\alpha}(\widetilde{\theta}  - \sigma) - \lambda(S,\sigma, t)]V_{\sigma} -rV = 0 .
 \end{equation}
Here, $\lambda (S,\sigma ,t) $ is the market price of volatility
risk which is usually assumed to be proportional to the variance:
$\lambda (S,\sigma,t) = \lambda_0 \sigma(t) $, where $\lambda_0 \in
\mathbb{R}$. The boundary conditions and final condition are
determined by the type of financial derivative $V(S,\sigma,t)$ we are
solving for. For example, in the case of the European Put Option:
\begin{align*}
    V(S,\sigma, T) &= \max(E-S,0),  & 0<&S<\infty, \; 0<\sigma<\infty,\\
    \lim_{S\to\infty}V(S,\sigma, t) &= 0, & 0<&\sigma<\infty, \; 0< t<T,\\
    V(0,\sigma, t) &= E\exp(-r(T-t)), & 0<&\sigma<\infty, \; 0< t<T,\\
    \lim_{\sigma\to\infty}V_{\sigma}(S,\sigma, t) &= 0, & 0<&S<\infty,\; 0< t<T,\\
\end{align*}
 The remaining boundary condition at $\sigma=0$ can be obtained by looking at
 the formal limit $\sigma\to 0$ in \eqref{PDE}, i.e.,
 \begin{equation}
   V_t+rSV_S+\kappa^*\theta^* V_\sigma-rV= 0,\quad T> t\geq 0,\;S>0,\; \text{as } \sigma\to  0.
   \label{boundary3}
 \end{equation}
 This boundary condition is used frequently, e.g.\ in \cite{IkoToi07,ZvFoVe98}.
 Alternatively, one can use a homogeneous Neumann condition
 \cite{ClaPar99}, i.e.,
 \begin{equation}
     V_{\sigma}(S,0,t) = 0, \quad 0<S<\infty, \; 0 < t<T.
 \end{equation}
\medskip

\noindent By using a change of variables:
$$ x=\ln\frac{S}{E},\hskip10pt y = \frac{\sigma}{v}, \hskip10pt \tau=T-t, \hskip8pt u = \exp(r\tau)\frac{V}{E}, \hskip10pt \kappa = \widetilde{\kappa}+\lambda_{0}, \hskip10pt \theta = \frac{\widetilde{\kappa}\widetilde{\theta}}{\widetilde{\kappa}+\lambda_{0}},    $$
we transform the partial differential equation to an convection-diffusion equation in two spatial dimensions with a mixed derivative term. The transformed partial differential equation and boundary/initial conditions are now satisfied by $u(x,y,\tau)$, where $x \in \mathbb{R}$, $y >0$, $\tau \in (0,T]$:
\begin{equation} \label{PDEtransf}
u_{\tau}=\frac{vy}{2}u_{xx}+\frac{(vy)^{2\beta}}{2}u_{yy} + \rho (v
y)^{\beta+\frac{1}{2}} u _{xy} + \Big(r-\frac{vy}{2}\Big)u_{x} +
\kappa(vy)^{\alpha}\frac{\theta - vy}{v}u_y,
 \end{equation}
\begin{subequations}
\begin{align}
   u(x,y,0) &= \max(1-\exp(x),0), & -\infty<&x<\infty, 0<y<\infty, \\
    \lim_{x\to\infty}u(x,y,\tau) &= 0 , &  0<&y<\infty, 0\leq \tau<T ,\\
    \lim_{x\to -\infty}u(x,y,\tau) &= 1 , &   0<&y<\infty, 0\leq
                                                  \tau<T ,\\
\label{bcymax}
    \lim_{y\to \infty}u_{y}(x,y,\tau)&=0 , & -\infty<&x<\infty, 0<
                                                       \tau\leq T,\\
\label{bcymin}
    \lim_{y\to 0}u_{y}(x,y,\tau)&=0 , &   -\infty<&x<\infty, 0< \tau\leq T.
\end{align}
\end{subequations}
In order to discretise the problem and solve numerically, we truncate our spatial boundaries to finite values. Take $ L_{1} \leq x \leq K_{1} $, where $ L_{1} < K_{1}$, and $ L_{2} \leq y \leq K_{2} $, where $0<L_2<K_{2}$, so that the spatial domain forms a closed rectangle in $\mathbb{R}^2$ of $M \times N$ points with uniform spacing of $\Delta_{x}$ in the $x$-direction and $\Delta_{y}$ in the $y$-direction:
$$x_{i} = L_{1} +(i-1)\Delta_{x},\; i=1,2,\ldots, M, \hspace{20pt} y_{j} = L_2 + (j-1)\Delta_{y},\;  j=1,2,\ldots, N.$$
The lower $y$-boundary is truncated to $ L_{2}> 0$ to ensure
non-degeneracy of the partial differential equation for all values of $y$.  We also take a uniform partition of $\tau\in [0,T]$ into $P$ points such that $\tau_{k} =  (k-1)\Delta_{\tau}$, where $k = 1,2,\ldots,P$. We denote the discrete approximation of $u((i-1)\Delta_x,(j-1)\Delta_y,(k-1)\Delta_\tau)$ by $u^{k}_{i,j}$ and $U^n=(u_{i,j}^n)_{i,j}$.

\section{Hundsdorfer-Verwer ADI splitting scheme}
\label{sec:HV}

\noindent We consider the Alternating Direction Implicit (ADI)
time-stepping numerical method proposed by Hundsdorfer and Verwer
\cite{Hund02, Verwer}. Our partial differential equation \eqref{PDEtransf} 
takes the form $u_{\tau} =F(u)$. We employ the splitting $F(u) =
F_{0}(u)+F_{1}(u)+F_{2}(u) $  where unidirectional and mixed
derivative differential operators are given by:
\begin{equation}\label{splitting}
F_{0}(u) = \rho (v y)^{\beta+\frac{1}{2}} u _{xy}, \; F_{1}(u) =
\frac{vy}{2}u_{xx}+\Big(r-\frac{vy}{2}\Big)u_{x}, \; F_{2}(u)=
\frac{(vy)^{2\beta}}{2}u_{yy} + \kappa(vy)^{\alpha}\frac{\theta -
  vy}{v}u_y .
\end{equation}
We consider \eqref{PDEtransf} with the splitting \eqref{splitting} 
and look for a semi-discrete approximation $U^n
\approx u(\tau_n)$  at time $n\Delta_{\tau}$.
Given an approximation $U^{n-1}$we
can calculate an approximation for $U^{n}$ at time
$n\Delta_{\tau}$ using the differential operators from
\eqref{splitting}:
\begin{subequations}
\label{HVscheme}
\begin{align}
Y_0 &=  U^{n-1}+\Delta_{t}F(U^{n-1}),  \\
Y_1 &=   Y_0 + \phi \Delta_{t}(F_{1}(Y_1)-F_{1}(U^{n-1})), \\
Y_2 &=  Y_1 + \phi \Delta_{t}(F_{2}(Y_2)-F_{2}(U^{n-1})),\\
\widetilde{Y}_0 &= Y_0 + \psi \Delta_{t}(F(Y_2)-F(U^{n-1})), \\
\widetilde{Y}_1 &=  \widetilde{Y}_0+\phi \Delta_{t}(F_{1}(\widetilde{Y}_1)-F_{1}(Y_2)),\\
\widetilde{Y}_2 & =  \widetilde{Y}_1+\phi \Delta_{t}(F_{2}(\widetilde{Y}_2)-F_{2}
(Y_2)),\\
U^n &= \widetilde{Y}_2.
\end{align}
\end{subequations}
The parameter $\psi$ is taken to be $\psi=1/2$ to ensure
second-order accuracy in time. The choice of $\phi$ is discussed in
\cite{Hund02}. Typically it is fixed to $\phi=1/2$. Larger values give
stronger damping of the implicit terms while lower values return better accuracy.

The first and fourth step in \eqref{HVscheme} can be solved explicitly, while the
remaining steps are solved implicitly. Our aim is to derive high-order
spatial discretisations of the differential operators. Following
\cite{DuFoRi13} we combine high-order compact finite difference
methods for the implicit steps with a (classical, non-compact)
high-order stencil for the explicit steps.


\section{High-order compact scheme for implicit steps}
\label{sec:HOC}

\noindent For $F_{1}(u)$, consider
\begin{equation}\label{F1}
F_{1}(u) = \frac{vy}{2}u_{xx}+\Big(r-\frac{vy}{2}\Big)u_{x} = g 
\end{equation}
with arbitrary right hand side $g$. We wish to derive a fourth-order
accurate in space approximation for \eqref{F1} which can be used to solve
the implicit second and fifth step in \eqref{HVscheme}. 
 Using standard second-order central difference operators and Taylor's expansion, we have:
\begin{align}
\label{ux}
 u_x(x_i,y_j) &= \delta_{x0} u_{i,j}
                -\frac{\Delta_x^2}{6}u_{xxx}(x_i,y_j)+\mathcal{O}(\Delta_x^4)
  \\ 
\label{uxx}
 u_{xx}(x_i,y_j) &= \delta_x^2 u_{i,j} -\frac{\Delta_x^2}{12}u_{xxxx}(x_i,y_j)+\mathcal{O}(\Delta_x^4) 
\end{align}
where 
$$\delta_{x0} u_{i,j}=\frac{u_{i+1,j} -  u_{i-1,j}}{2\Delta_x}\text{
  and } \delta_x^2 u_{i,j} = \frac{u_{i+1,j} - 2u_{i,j} +
  u_{i-1,j}}{\Delta^{2}_x}.$$ 
If we can find second-order accurate expressions for $u_{xxx}$ and
$u_{xxxx}$ using only information on the compact stencil, then it will
be possible to approximate $u_x$ and $u_{xx}$ with fourth order
accuracy on the compact stencil. By differentiating \eqref{F1}  once
and twice with respect to $x$, respectively, it is possible to express $u_{xxx}$ and $u_{xxxx}$ in terms of first- and second-order derivatives of $u$ and $g$ with respect to $x$:
\begin{align}
\label{uxxx}
u_{xxx}&= \frac{2}{vy}g_x+\Big(1-\frac{2r}{vy}\Big)u_{xx},\\ 
\label{uxxxx}
u_{xxxx} &= \frac{2}{vy}g_{xx} +\Big(1-\frac{2r}{vy}\Big)\Big[\frac{2}{vy}g_x+\Big(1-\frac{2r}{vy}\Big)u_{xx}\Big].
\end{align}

By substituting standard second-order central difference operators into $\eqref{uxxx}$ and $\eqref{uxxxx}$ we obtain second-order accurate in space approximations for $u_{xxx}$ and $u_{xxxx}$:
\begin{align}
\label{uxxxdiscr}
u_{xxx}(x_i,y_j) &= \frac{2}{vy_j}\delta_{x0}
                   (g_{i,j})_x+\Big(1-\frac{2r}{vy_j}\Big)\delta_x^2
                   u_{i,j}+O(\Delta_x^2), \\ 
\label{uxxxxdiscr}
u_{xxxx}(x_i,y_j) &= \frac{2}{vy_j}\delta_x^2 g_{i,j} +\Big(1-\frac{2r}{vy_j}\Big)\Big[\frac{2}{vy_j}\delta_{x0} g_{i,j}+\Big(1-\frac{2r}{vy_j}\Big)\delta_x^2 u_{i,j}\Big]+\mathcal{O}(\Delta_x^2).
\end{align}
Substituting \eqref{uxxxdiscr} and \eqref{uxxxxdiscr} into \eqref{ux} and \eqref{uxx}, respectively, yields:
\begin{align}
 u_x(x_i,y_j) &= \delta_{x0} u_{i,j} -\frac{\Delta_x^2}{6}\Big[\frac{2}{vy_j}\delta_{x0} (g_{i,j})_x+\Big(1-\frac{2r}{vy_j}\Big)\delta_x^2 u_{i,j}\Big]+\mathcal{O}(\Delta_x^4), \\ 
 u_{xx}(x_i,y_j) &= \delta_x^2 u_{i,j} -\frac{\Delta_x^2}{12}\Bigg[\frac{2}{vy_j}\delta_x^2 g_{i,j} +\Big(1-\frac{2r}{vy_j}\Big)\Big[\frac{2}{vy_j}\delta_{x0} g_{i,j}+\Big(1-\frac{2r}{vy_j}\Big)\delta_x^2 u_{i,j}\Big]\Bigg]+\mathcal{O}(\Delta_x^4 ).
\end{align}
Substituting these fourth-order approximations for $u_x$ and $u_{xx}$
into \eqref{F1} and rearranging the equation such that all derivatives of $u$ with respect to $x$ are on the left hand side and all derivatives of $g$ with respect to $x$ are on the right hand side we obtain a fourth-order compact scheme for \eqref{F1}:
\begin{multline}
\label{F1HOC}
\Big(\frac{vy_j}{2}-\frac{-v^2 y_j^2 \Delta_x^2+4rvy_j\Delta_x^2-4r^2\Delta_x^2}{24vy_j}\Big)\delta_x^2 u_{i,j}+\Big(r-\frac{vy_j}{2}\Big)\delta_{x0} u_{i,j} \\  = g_{i,j}
+ \frac{-2vy_j\Delta_x^2+4r\Delta_x^2}{24vy_j}\delta_{x0} g_{i,j}+\frac{\Delta_x^2}{12}\delta_x^2 g_{i,j} .
\end{multline}
Finally, substituting the expressions for the difference operators
$\delta_{x0}$, $\delta_x^2$ into \eqref{F1HOC} and separating the terms into values of $u$ and $g$ at the three horizontally adjacent nodal points in space, we get:
\begin{multline}
\label{F1HOCnodal}
\frac{v^2y_j^2\Delta_x^2-4rvy_j\Delta_x^2-6v^2y_j^2\Delta_x+4r^2\Delta_x^2+12rvy_j\Delta_x+12v^2y_j^2}{24vy_j\Delta_x^2}{\it
    u_{i+1,j}} \\ -
  \frac{v^2y_j^2\Delta_x^2-4rvy_j\Delta_x^2+4r^2\Delta_x^2+12v^2y_j^2}{12vy_j\Delta_x^2}{\it
    u_{i,j}} \\+
  \frac{v^2y_j^2\Delta_x^2-4rvy_j\Delta_x^2+6v^2y_j^2\delta_x+4r^2\Delta_x^2-12rvy_j\Delta_x+12v^2y_j^2}{24vy_j\Delta_x^2}{\it
    u_{i-1,j}}\\ =  \frac{-vy_j\Delta_x+2r\Delta_x+2vy_j}{24vy_j}{\it
    g_{i+1,j}}+\frac{5}{6}{\it
    g_{i,j}}-\frac{-vy_j\Delta_x+2r\Delta_x-2vy_j}{24vy_j}{\it g_{i-1,j}}
\end{multline}
Equation \eqref{F1HOCnodal} defines a fourth-order compact
approximation for \eqref{F1HOCnodal}. In other words, we have a system of equations
which defines a fourth-order accurate approximation for \eqref{F1HOCnodal} at any
point on the inner grid of the spatial domain (all points of the
spatial domain except those that lie on the $x$ and $y$
boundaries). To approximate \eqref{F1HOCnodal} at points along the $x$ boundaries
of the inner grid of the spatial domain, we will require a
contribution from the Dirichlet values at the $x$-boundaries of the
spatial domain. We collect these separately in a vector
$d$. Details on the boundary conditions are given in Section~\ref{sec:bc}.
The linear system to be solved can be written in matrix form:
$$A_{x}{{u}} = B_{x}{{g}} + d,$$
where ${{u}} = (u_{2,2}, u_{2,3}, \ldots, u_{N-1,M-1})$,
${{g}} = (g_{2,2}, g_{2,3}, \ldots, g_{N-1,M-1})$. The coefficient matrices $A_x$ and $B_x$ are block diagonal matrices, with the following structure:
\begin{equation*} {A_x} = \left[ {\begin{array}{cccc}
A_x^{1,1}& 0 & 0  & 0\\
0 & A_x^{2,2} & 0 & 0\\
0 & 0 & \ddots & 0  \\
0 &0 &0 &A_x^{N-2,N-2} 
\end{array}} \right],\quad 
{B_x} = \left[ {\begin{array}{cccc}
B_x^{1,1}& 0 & 0  & 0\\
0 & B_x^{2,2} & 0 & 0\\
0 & 0 & \ddots & 0  \\
0 &0 &0 &B_x^{N-2,N-2} 
\end{array}} \right],
\end{equation*} 
where each $A_x^{j,j} = \mathrm{diag}[a_{-1}^{j,j},a_0^{j,j},a_1^{j,j}]$ and $B_x^{j,j} = \mathrm{diag}[b_{-1}^{j,j},b_0^{j,j},b_1^{j,j}]$ are tri-diagonal matrices.
\medskip

Let us consider now the case of $F_{2}$:
\begin{equation}
\label{F2}
 F_{2} (u)=  \frac{(vy)^{2\beta}}{2}u_{yy} + \kappa(vy)^{\alpha}\frac{\theta - vy}{v}u_y=g .
\end{equation} 
Due to the appearance of $y$ terms in the coefficients of $F_2(u)$,
the algebraic complexity in deriving a fourth-order accurate scheme in
space is much greater. By Taylor's expansions we obtain:
\begin{align}
\label{uy}
u_y\left( x_{i},y_{j} \right) 
=\delta _{y_{0}}u_{i,j}-\frac{{\Delta_y}^{2
}}{6}u_{yyy} \left( x_{i},y_{j
} \right) +\mathcal{O} (\Delta_y^{4}) , \\
\label{uyy}
u_{yy}\left( x_{{i}},y_{{j}
} \right) ={\delta^{2}_{y}}u_{{i,j}}-\frac{{\Delta_y}^{2}}{12}u_{yyyy} \left( x
_{{i}},y_{{j}} \right) +\mathcal{O} (\Delta_y^{4}).
\end{align}
We wish to find second order accurate approximations for $u_{yyy}$ and
$u_{yyyy}$ on the compact stencil in order to find fourth-order
accurate expressions for $u_y$ and $u_{yy}$. Re-arranging \eqref{F2},
we get:
$$
u_{yy} = \frac{2}{(vy)^{2\beta}}\Big(-\kappa(vy)^{\alpha}\frac{(\theta - vy)}{v}u_y+g\Big).
$$
Via repeated applications of the chain rule, second-order accurate approximations for $u_{yyy}(x_i,y_j)$ and $u_{yyyy}(x_i,y_j)$ are given by:
\begin{multline} 
\label{uyyy}
u_{yyy}(x_i,y_j) = \frac 
{\left( 2 \left( vy_j \right) ^{\alpha}\alpha kvy_j-2 \left( vy_j
 \right) ^{\alpha}\theta\alpha k+2 \left( vy_j \right) ^{\alpha}kvy_j
 \right)  }{
  \left( vy_j \right) ^{2\beta} vy_j} \delta_{y0}u_{i,j} \\ 
 +\frac { ( 2 \left( vy_j \right) ^{\alpha} kv {y_j}^{2} - 2   \left( vy_j \right)^{2\beta} \beta v-2 \left( vy_j \right) ^{\alpha}\theta k y_j ) } {\left(vy_j\right)^{2\beta}vy_j} \delta^2_y u_{i,j}  +\frac {2}{ \left(vy_j\right)^{2\beta}}\delta_{y0}g_{i,j}+\mathcal{O}(\Delta^2_y),
\end{multline}
\begin{multline}
\label{uyyyy}
u_{yyyy} ( x_i,y_j) = \left( {\frac {2(2 \left( vy_j \right) ^{\alpha}kv{y_j}^{2}-2
 \left(vy_j\right)^{2\beta}\beta v-2 \left( vy_j
 \right) ^{\alpha}\theta ky_j)}{\left(vy_j\right)^{4\beta}vy_j}}-{\frac {4\beta}{ \left(vy_j\right)^{2\beta}y_j}} \right) \delta_{y0}g_{i,j} \\ + \Bigg( {\frac {1}{ \left(vy_j\right)^{2\beta}vy_j} ( 2 \left( vy_j \right) ^{\alpha}{\alpha}^{2}kv+
4 \left( vy_j \right) ^{\alpha}\alpha kv-{\frac { 2\left( vy_j
 \right) ^{\alpha}{\alpha}^{2}\theta k}{y_j}}+2 \left( vy_j \right) ^{
\alpha}kv ) } \\ -{\frac {2 \beta \left( 2 \left( vy_j \right) ^{\alpha}
\alpha kvy_j-2 \left( vy_j \right) ^{\alpha}\theta \alpha k+2
 \left( vy_j \right) ^{\alpha}kvy_j \right) }{\left(vy_j\right)^{2\beta}v{y_j}^{2}}}-{\frac {2 \left( vy_j
 \right) ^{\alpha}\alpha kvy_j-2 \left( vy_j \right) ^{\alpha}\theta
\alpha k+2  \left( vy_j \right) ^{\alpha}kvy_j}{\left(vy_j\right)^{2\beta}v{y_j}^{2}}} \\ +{\frac { ( 2 \left( vy_j
 \right) ^{\alpha}kv{y_j}^{2}-2\left(vy_j\right)^{2\beta}\beta v-2  \left( vy_j \right) ^{\alpha}\theta ky_j
 )  \left( 2 \left( vy_j \right) ^{\alpha}\alpha kvy_j-2
 \left( vy_j \right) ^{\alpha}\theta \alpha k+2 \left( vy_j \right) ^{
\alpha}kvy_j \right) }{ \left(vy_j\right)^{4\beta}{
v}^{2}{y_j}^{2}}} \Bigg)  \delta_{y0}u_{i,j} \\ + \Bigg( {\frac {2 \left( vy_j \right) ^{\alpha}\alpha kvy_j-2
\left( vy_j \right) ^{\alpha}\theta \alpha k+2 \left( vy_j \right) 
^{\alpha}kvy_j}{ \left(vy_j\right)^{2\beta}vy_j}} \\ + {
\frac {1}{ \left(vy_j\right)^{2\beta}vy_j} ( 2
 \left( vy_j \right) ^{\alpha}\alpha kvy_j+4 \left( vy_j \right) ^{
\alpha}kvy_j -4 {\frac { \left(vy_j\right)^{2\beta}
{\beta}^{2}v}{y_j}}-2 \left( vy_j \right) ^{\alpha}\theta \alpha k-2
 \left( vy_j \right) ^{\alpha}\theta k ) } \\ - {\frac {2\beta ( 2
 \left( vy_j \right) ^{\alpha}kv{y_j}^{2}-2 \left(vy_j\right)^{2\beta}\beta v-2 \left( vy_j \right) ^{\alpha}\theta ky_j
 )}{ \left(vy_j\right)^{2\beta}v{y_j}^{
2}}}-{\frac {2 \left( vy_j \right) ^{\alpha}kv{y_j}^{2}-2 \left(vy_j\right)^{2\beta}\beta v-2 \left( vy_j
 \right) ^{\alpha}\theta ky_j}{ \left(vy_j\right)^{2\beta}v{y_j}^{2}}} \\ + {\frac { ( 2 \left( vy_j \right) ^{\alpha
}kv{y_j}^{2}-2\left(vy_j\right)^{2\beta}\beta v
-2 \left( vy_j \right) ^{\alpha}\theta ky_j ) ^{2}}{ \left(vy_j\right)^{4\beta}{v}^{2}{y_j}^{2}}} \Bigg) \delta^2_y u_{i,j}+{
\frac {2}{ \left(vy_j\right)^{2\beta}}}\delta^2_y g_{i,j}+\mathcal{O}(\Delta^2_y).
\end{multline}
where $\delta_{y0}$ and $\delta_y^2$ denote the standard second-order
central difference operators.

Substituting \eqref{uyyy} and \eqref{uyyyy} into \eqref{uy} and
\eqref{uyy}, respectively, yields fourth-order accurate
approximations (not given here) for $u_y(x_i,y_j)$ and $u_{yy}(x_i,y_j)$ on the compact
stencil. By substituting these fourth-order accurate approximations
into \eqref{F2} and separating the $u$ and $g$ terms onto the left and
right hand sides, respectively, we obtain a linear system which can be represented in matrix form: $$A_{y}{{u}} = B_{y}{{g}}$$
where ${{u}} = (u_{2,2}, u_{2,3}, \ldots, u_{N-1,M-1})$,
${{g}} = (g_{2,2}, g_{2,3}, \ldots, g_{N-1,M-1})$. 
We do not impose any boundary conditions in $y$-direction, but
discretise the boundary grid points with the same scheme, and handle
resulting ghost points via extrapolation; details on the boundary
conditions are given in Section~\ref{sec:bc}. 
The coefficient matrices $A_y$ and $B_y$ are block tri-diagonal matrices with the following structures:
\begin{center}
\begin{eqnarray*}
\mathbf{A_y} = \left[\begin{array}{ccccc}
A_y^{1,1} & A_y^{1,2} & 0 & 0 & 0  \\
A_y^{2,1} & A_y^{2,2} & A_y^{2,3} & 0 & 0\\
0 & \ddots & \ddots & \ddots & 0  \\
0 & 0 & A_y^{N-3,N-4} & A_y^{N-3,N-3} & A_y^{N-3,N-2} \\ 0 & 0&0 & A_y^{N-2,N-3} & A_y^{N-2,N-2} 
\end{array}\right], \\ \mathbf{B_y} = \left[\begin{array}{ccccc}
B_y^{1,1} & B_y^{1,2} & 0 & 0 & 0  \\
B_y^{2,1} & B_y^{2,2} & B_y^{2,3} & 0 & 0\\
0 & \ddots & \ddots & \ddots & 0  \\
0 & 0 & B_y^{N-3,N-4} & B_y^{N-3,N-3} & B_y^{N-3,N-2} \\ 0 & 0&0 & B_y^{N-2,N-3} & B_y^{N-2,N-2} 
\end{array}\right],
\end{eqnarray*}
\end{center}
where each $A_y^{j,j} = \mathrm{diag}[a^{i,j}]$ and $B_y^{j,j} =
\mathrm{diag}[b^{i,j}]$ are diagonal matrices, with values on these diagonals given as follows:
\begin{multline}
a^{i,j\pm 1}=\frac{1}{2{\Delta_y}^2} (vy_j)^{2\beta}-\frac{1}{12(vy_j)^{2\beta}
{v}^{2}{y_j}^{2}}\Big(-2 (   vy_j )^{2\alpha} {\kappa}^{2}{v}^{2}{y_j}^{4} +2 (vy_j)^{2\beta+\alpha} \alpha \kappa{v}^{2}{y_j}^{2} \\
 -2 (vy_j)^{2\beta+\alpha}\beta \kappa{v}^{2}{y_j}^{2}+4   ( vy_j ) ^{2\alpha} \theta{k}^{2}v{y_j}^{3} +2  (vy_j)^{4\beta}{\beta}^{2}{v}^{2} -2 (vy_j)^{2\beta+\alpha} \theta\alpha \kappa v y_j \\ 
+2 (vy_j)^{2\beta+\alpha} \theta\beta \kappa v y_j  +2 (vy_j)^{2\beta+\alpha}\kappa{v}^{2}{y_j}^{2}  -2   ( vy_j ) ^{2\alpha} {\theta}^{2}{\kappa}^{2}{y_j}^{2}+ (vy_j)^{4\beta}\beta{v}^{2}\Big) \\ 
\pm\Bigg( \frac {- ( vy_j )^{\alpha}\kappa v y_j+ ( vy_j ) ^{\alpha}\theta \kappa}{2v\Delta_y} -\frac{1}{24\Delta_y\beta^2( vy_j)^4}\Big(-2   ( vy_j ) ^{2
\alpha}\alpha{\kappa}^{2}{v}^{2}{y_j}^{3}{\Delta_y}^{2}\\
+ (vy_j)^{2\beta} ( vy_j ) ^{\alpha}{\alpha}^{2}\kappa{v}^{2}y_j{\Delta_y}^{2} -4 (vy_j)^{2\beta+\alpha} \alpha\beta \kappa{v}^{2}y_j{\Delta_y}^{2}+4(vy_j)^{2\alpha}\theta\alpha{\kappa}^{2}v{y_j}^{2}{\Delta_y}^{2}\\
- 2  ( vy_j )^{2\alpha} {\kappa}^{2}{v}^{2}{y_j}^{3}{\Delta_y}^{2}-  ( vy_j )^{2\beta+\alpha} \theta{\alpha}^{2}\kappa v{\Delta_y}^{2}+4 ( vy_j )^{2\beta+\alpha}
\theta\alpha\beta \kappa v{\Delta_y}^{2}\\ 
+ ( vy_j )^{2\beta+\alpha}\alpha \kappa{v}^{2}y_j{\Delta_y}^{2} -4 ( vy_j )^{2\beta+\alpha}\beta
\kappa{v}^{2}y_j{\Delta_y}^{2}-2 ( vy_j )^{2\alpha}{\theta}^{2}\alpha{\kappa}^{2}y_j{\Delta_y}^{2}\\ 
+2 ( vy_j )^{2\alpha}\theta{\kappa}^{2}v{y_j}^{2}{\Delta_y}^{2}+ ( vy_j )^{2\beta+\alpha}\theta\alpha \kappa v{\Delta_y}^{2}\Big)\Bigg),
\end{multline}
\begin{multline}
a^{i,j}= \frac{1}{6( vy_j ) ^ {2\beta+2}}\Big( -2 ( vy_j ) ^ {2\alpha}{k}^{2}{v}^{2}{y_j}^{4}+2 ( vy_j ) ^ {2\beta+\alpha}\alpha k{v}^{2}{y_j}^{
2} +2 ( vy_j ) ^ {4\beta}{
\beta}^{2}{v}^{2}\\ -2 ( vy_j ) ^ {2\beta+\alpha}\beta k{v}^{2}{y_j}^{2}  +4
 ( vy_j ) ^ {2\alpha}\theta{k}^{2}v{y_j}^{
3} -2 ( vy_j ) ^ {2\beta+\alpha} \theta\alpha k v y_j + ( vy_j ) ^ {4\beta}\beta{v}^{2} \\   +2 ( vy_j ) ^ {2\beta+\alpha}\theta\beta k v y_j+2 ( vy_j ) ^ {2\beta+\alpha}k{v}^{2}{y_j
}^{2} - 2 ( vy_j ) ^ {2\alpha}{\theta}^{2}{k}^{2}{y_j}^{2} - 6
( vy_j ) ^ {4\beta+2} )\Big),
\end{multline}
\begin{equation}
b^{i,j\pm 1}= \pm{\frac {-2 ( vy_j ) ^{\alpha}k{v}^{2}{y_j}^{3}{
\Delta_y}^{2}-4 ( vy_j ) ^ {2\beta}\beta
{v}^{2}y_j{\Delta_y}^{2}+2 ( vy_j ) ^{\alpha}\theta kv{y_j}^
{2}{\Delta_y}^{2}}{ 24( vy_j ) ^ {2\beta+2}\Delta_y}}+\frac{1}{12},\quad b^{i,j}=\frac{5}{6}.
\end{equation}

\section{High-order scheme for explicit steps}
\label{sec:HOexpl}

\noindent The first and fourth steps of the ADI scheme
\eqref{HVscheme} operate only on previous approximations to explicitly 
calculate an updated approximation. The differential operator in these
steps takes the form of the right hand side of \eqref{PDEtransf}. For
the mixed derivative term it seems not to be possible to
exploit the structure of the differential operator to
obtain a fourth-order approximation on a compact computational
stencil. Hence, in order to maintain fourth-order accuracy of the
scheme in the explicit steps of \eqref{HVscheme}, the derivatives in
each differential operator  $F_0$, $F_1$ and $F_2$ are approximated
using classical, fourth-order central difference operators which
operate on a larger $5\times 5$-stencil in the spatial domain. 

For $F_{1}(u) = \frac{vy}{2}u_{xx}-(\frac{vy}{2}-r)u_{x}$, we have the following scheme: 
\begin{multline*}
\Big[ \frac{vy}{2} \frac{\partial^2 u}{\partial x^2}+\Big (r-\frac{vy}{2}\Big)\frac{\partial u}{\partial x}\Big]_{i,j} = \Big (\frac{2r-vy_j}{24\Delta_x}-\frac{vy_j}{24\Delta_x^2}\Big)u_{i,j-2} + \Big (\frac{8vy_j-16r}{24\Delta_x}+\frac{16vy_j}{24\Delta_x^2}\Big)u_{i,j-1} \\ - \frac{30vy_j}{24\Delta_x^2}u_{i,j} + \Big (\frac{16r-8vy_j}{24\Delta_x}+\frac{16vy_j}{24\Delta_x^2}\Big)u_{i,j+1} + \Big (\frac{vy_j-2r}{24\Delta_x}-\frac{vy_j}{24\Delta_x^2}\Big)u_{i,j+2} + \mathcal{O}(\Delta^4_x).
\end{multline*}
For $F_{2}(u)=\frac{(vy)^{2\beta}}{2}u_{yy}+\frac{\kappa(vy)^{\alpha}(\theta-vy)}{v}u_{y}$, we have:
\begin{multline*}
\Big[\frac{(vy)^{2\beta}}{2}\frac{\partial^2 u}{\partial y^2}+\frac{\kappa(vy)^{\alpha}(\theta-vy)}{v}\frac{\partial u}{\partial y}\Big]_{i,j} = \Big (-\frac{\kappa(vy_j)^{\alpha}(\theta-vy_j)}{12v\Delta_y}-\frac{(vy_j)^{2\beta}}{24\Delta_y^2}\Big)u_{i-2,j} \\
+ \Big (\frac{8\kappa(vy_j)^{\alpha}(\theta-vy_j)}{12v\Delta_y}+\frac{16(vy_j)^{2\beta}}{24\Delta_y^2}\Big)u_{i-1,j}  - \frac{30(vy_j)^{2\beta}}{24\Delta_y^2}u_{i,j} \\
+ \Big (-\frac{8\kappa(vy_j)^{\alpha}(\theta-vy_j)}{12v\Delta_y}+\frac{(vy_j)^{2\beta}}{24\Delta_y^2}\Big)u_{i+1,j} + \Big (\frac{\kappa(vy_j)^{\alpha}(\theta-vy_j)}{12v\Delta_y}-\frac{(vy_j)^{2\beta}}{24\Delta_y^2}\Big)u_{i+2,j} + \mathcal{O}(\Delta^4_y ).
\end{multline*}
Finally, for the mixed derivative term $F_0 = \rho (v y)^{\beta+\frac{1}{2}}u_{xy} $, the following computational stencil is used:
\begin{multline*}
    \Big[\rho (v y)^{\beta+\frac{1}{2}} \frac{\partial^2 u}{\partial x \partial
      y}\Big]_{i,j} = -\frac{64\rho (v y_j)^{\beta+\frac{1}{2}}}{144\Delta_x\Delta_y}u_{i-1,j-1}
      + \frac{64\rho (v y_j)^{\beta+\frac{1}{2}}}{144\Delta_x\Delta_y}u_{i-1,j+1} 
      + \frac{64\rho (v y_j)^{\beta+\frac{1}{2}}}{144\Delta_x\Delta_y}u_{i+1,j-1}  \\
-  \frac{64\rho (v y_j)^{\beta+\frac{1}{2}}}{144\Delta_x\Delta_y}u_{i+1,j+1} 
- \frac{\rho (v y_j)^{\beta+\frac{1}{2}}}{144\Delta_x\Delta_y}u_{i-2,j-2}
+ \frac{8 \rho (v y_j)^{\beta+\frac{1}{2}}}{144\Delta_x\Delta_y}u_{i-2,j-1} 
- \frac{8 \rho (v y_j)^{\beta+\frac{1}{2}}}{144\Delta_x\Delta_y}u_{i-2,j+1}\\
+ \frac{ \rho (v y_j)^{\beta+\frac{1}{2}}}{144 \Delta_x\Delta_y}u_{i-2,j+2} + 
\frac{8 \rho (v y_j)^{\beta+\frac{1}{2}}}{144 \Delta_x\Delta_y}u_{i-1,j-2} 
- \frac{8 \rho (v y_j)^{\beta+\frac{1}{2}}}{144\Delta_x\Delta_y}u_{i-1,j+2} 
- \frac{8 \rho (v y_j)^{\beta+\frac{1}{2}}}{144\Delta_x\Delta_y}u_{i+1,j-2}  \\ 
+\frac{8 \rho (v y_j)^{\beta+\frac{1}{2}}}{144 \Delta_x\Delta_y}u_{i+1,j+2} 
+ \frac{ \rho (v y_j)^{\beta+\frac{1}{2}}}{144 \Delta_x\Delta_y}u_{i+2,j-2} 
-\frac{8 \rho (v y_j)^{\beta+\frac{1}{2}}}{144\Delta_x\Delta_y}u_{i+2,j-1} 
+\frac{8 \rho (v y_j)^{\beta+\frac{1}{2}}}{144\Delta_x\Delta_y}u_{i+2,j+1} \\
- \frac{ \rho (v y_j)^{\beta+\frac{1}{2}}}{144 \Delta_x\Delta_y}u_{i+2,j+2} + 
\mathcal{O}(\Delta^2_x\Delta^2_y).
\end{multline*}

Using these fourth-order approximations, the first and fourth step in
\eqref{HVscheme} can be computed directly. The values at the spatial
boundaries for each solution of the ADI scheme are determined by the
boundary conditions, the computational stencil is required for all
remaining points in the spatial domain. For the explicit steps, the
$5\times 5$-point computational stencil exceeds the spatial boundary
when we wish to approximate differential operator $F(u)$ at any point
along the boundary of the spatial domain's inner grid. For example if
we wish to evaluate $F(u_{2,2})$, we will require contributions from
ghost points which fall outside the spatial domain, as marked by
bullet points in Figure~\ref{fig:ghostpoints}.
\begin{figure}
$$\begin{array}{c|ccccc}
\bullet & \mathrm{u_{4,1}} & u_{4,2} & u_{4,3} & u_{4,4} \\
\bullet & \mathrm{u_{3,1}} & u_{3,2} & u_{3,3} & u_{3,4} \\
\bullet & \mathrm{u_{2,1}} & {{u_{2,2}}} & u_{2,3} & u_{2,4} \\
\bullet & \mathrm{u_{1,1}} & \mathrm{u_{1,2}} & \mathrm{u_{1,3}} & \mathrm{u_{1,4}} \\
\hline 
\odot & \circ & \circ & \circ & \circ 
\end{array}$$
\caption{Example: evaluation of $F(u_{2,2})$ using the $5\times 5$-point computational stencil in the lower left
  corner of the computational domain; ghost points outside the
  computational domain at which values are
extrapolated from the interior of the domain are marked by bullets ($\bullet$,$\circ$,$\odot$),
grid points on the boundary are set in Roman.}
\label{fig:ghostpoints}
\end{figure}
We extrapolate information from grid points $u(x_i,y_j)$, where $i =
1,\ldots ,M-1,$ $j = 1,\ldots, N-1$ to establish values at these ghost
points for the purpose of evaluating the differential operator $F(u)$ at any point along the boundary of the inner grid of the spatial domain. To calculate the values at these ghost points, we use the following five-point extrapolation formulae for three cases:
\begin{align*}
& x=L_1\text{ boundary }(\bullet)\text{ :} &u_{i,0} &= 5u_{i,1} - 10u_{i,2} + 10 u_{i,3} -5u_{i,4} + u_{i,5} + \mathcal{O}(\Delta^6_x) ,\\
& y=L_2\text{ boundary }(\circ)\text{ :} &u_{0,j} &=  5u_{1,j} - 10u_{2,j} + 10 u_{3,j} -5u_{4,j} + u_{5,j} + \mathcal{O}(\Delta^6_y),\\
& x=L_1, y=L_2\text{ corner }(\odot)\text{ :} &u_{0,0} &=  5u_{1,1} - 10u_{2,2}+ 10 u_{3,3} -5u_{4,4} + u_{5,5} +\mathcal{O}(\Delta^3_x\Delta^3_y).
\end{align*}
The extrapolation at the $x=K_1$ and $y=K_2$ boundaries
and the remaining three corners is handled analogously. 

\section{Solving the high-order ADI scheme}
\label{sec:HOsolve}

\noindent Starting from a given $U^{n-1}$, the ADI
scheme \eqref{HVscheme} involves six approximation steps to obtain $U^{n}$, the
solution at the next time level. The first approximation $Y_0$ can be
solved for explicitly using the $5\times 5$-point computational
stencil derived in Section~\ref{sec:HOexpl}. The second approximation
for our solution, denoted by $Y_1$, has to be solved for implicitly:
\begin{align}
\label{stepximpl}
Y_1 =& Y_0 + \phi{\Delta_{t}} (F_{1}(Y_1)-F_{1}(U^{n-1}))\quad
\Longleftrightarrow \quad F_1(Y_1-U^{n-1})= \frac1{\phi\Delta_t}(Y_1-Y_0).
\end{align}
We apply the fourth-order compact scheme established in
Section~\ref{sec:HOC} to solve \eqref{stepximpl}. In matrix form we obtain
 $$A_x(Y_1-U^{n-1}) = B_x\Big(\frac{1}{\phi\Delta_t}(Y_1-Y_0)\Big)+ d.$$
Collecting unknown $Y_1$ terms on the left hand side and known terms
$Y_0$, $U^{n-1}$ and $d$ on the right hand side we get
$$\left(B_x-\phi\Delta_t A_x\right)Y_1 = B_{x}Y_0 - \phi\Delta_tA_{x}U^{n-1}- \phi\Delta_td.$$
To solve, we invert the tri-diagonal matrix
$\left(B_x-\phi\Delta_t A_x\right)$. 
For the third step of the ADI scheme, we proceed analogously, and use
the the high-order compact scheme presented in Section~\ref{sec:HOC}
to solve for $Y_2$ implicitly. The fourth, fifth and sixth step of the
ADI scheme are performed analogously as the first, second
and third steps, respectively.  

Note that the matrix $\left(B_x-\phi\Delta_t A_x\right)$ appears
twice in the scheme \eqref{HVscheme}, in the second and fifth step. Similarly,
$\left(B_y-\phi\Delta_tA_y\right)$ appears in the third and the sixth step. Hence, using LU-factorisation, only two matrix inversions are
necessary in each time step of scheme \eqref{HVscheme}.
Moreover, since the coefficients in the partial differential equation
\eqref{PDEtransf} do not depend on time, and the matrices are therefore
constant, they can be LU-factorised before iterating in time to obtain
a highly efficient algorithm.

The combination of the fourth-order spatial discretisation presented in
Section~\ref{sec:HOC} and \ref{sec:HOexpl} with the second-order time
splitting \eqref{HVscheme} yields a high-order ADI scheme with order of consistency two in time and four in space.

\section{Boundary conditions}
\label{sec:bc}

\noindent For the case of the Dirichlet conditions at $x=L_1$ and
$x=K_1$ we impose 
\begin{align*}
 u(L_1,y_j,\tau_k)&= 1-e^{r\tau+L_1}, & j&=1,2,\ldots,N,\;k=1,2,\ldots
, \\
u(K_1,y_j,\tau_k)&= 0, &  j&=1,2,\ldots,N,\;k=1,2,\ldots.
\end{align*}
Using the homogeneous Neumann conditions \eqref{bcymax} and \eqref{bcymin} which are correct in
the limit $y\to \infty$ and $y\to 0$, respectively, at the (finite)
boundaries $y=L_2>0$ and $y=K_2$ would result in a dominant error along
these boundaries. Hence, we do not impose any boundary condition at
these two boundaries but discretise the partial derivative using the computational stencil
from the interior.
The values of the unknown on the boundaries are set by extrapolation
from values in the interior. This introduces a numerical error, and it
needs to be considered that the order of extrapolation should be high
enough not to affect the overall order of accuracy. We refer to Gustafsson \cite{GusBC} to discuss the influence of the  
order of the approximation on the global convergence rate.
We use the following extrapolation formulae:
\begin{align*}
& u^k_{i,1} = 5u^k_{i,2} -10u^k_{i,3} +10u^k_{i,4} -5u^k_{i,5}+u^k_{i,6} +\mathcal{O}(\Delta^6_y),\\ & u^k_{i,N} = 5u^k_{i,N-1} -10u^k_{i,N-2} +10u^k_{i,N-3} -5u^k_{i,N-4} +u^k_{i,N-5} +\mathcal{O}(\Delta^6_y).
\end{align*}

\section{Numerical experiments}
\label{sec:num}

\noindent In this section we report the results of our numerical
experiments. We estimate the numerical convergence order of the
high-order ADI scheme and then
perform additional experiments to validate its stability.

\subsection{Numerical convergence}

\noindent We perform a numerical study to compute the order of
convergence of the high-order ADI scheme. 
Since the initial condition for the option pricing problem,
the payoff function $V(S,\sigma, T)$, is non-smooth at $S=E$, we
cannot in general expect to observe high-order convergence \cite{KrThWi70}.
A straightforward way to smooth the initial condition is to choose the
mesh in such a way that the non-smooth point of the initial condition is
not a point of the mesh.  The construction of such a mesh is always
possible in a simple manner. Following this approach, the non-smooth payoff can be directly considered in our
scheme and, indeed,  we observe high-order numerical convergence. 
Alternatively, suitable smoothing operators can be employed to
achieve a similar effect, see \cite{KrThWi70,DuHe15}.

\begin{figure}
\centering
\includegraphics[width=0.75\textwidth]{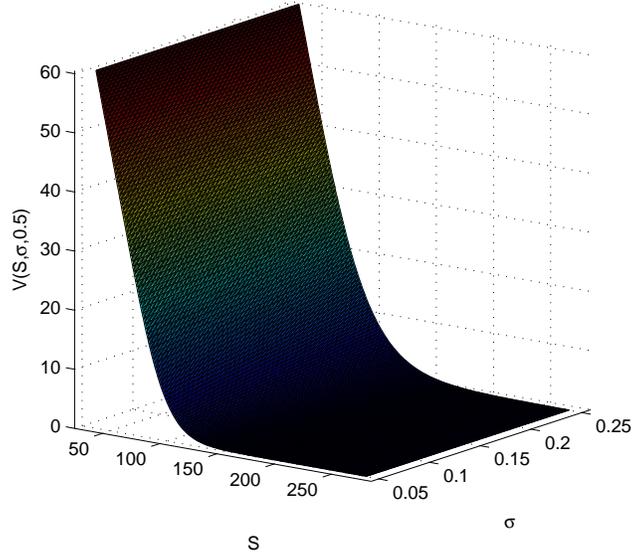}
\caption{Numerical solution for price of European Put Option at $T=0.5$}
\label{fig:1}
\end{figure}

For convenience, we choose an equally sized space step $h = \Delta_x =
\Delta_y,$ creating an  evenly-spaced mesh both horizontally and
vertically. We set the parameter $\phi=0.5$ in \eqref{HVscheme} in all
numerical experiments. 
Figure~\ref{fig:1} shows the numerical
solution for the European option price at time $T=0.5$ using the
parameters from Table~\ref{defaultparams}.
\begin{table}
\begin{center}
 \begin{tabular}[c]{l c}
\toprule
 Parameter & Value  \\ 
\midrule
 Strike price & $E=100$  \\ 
 Time to maturity & $T = 0.5$ \\
Interest rate      &  $r=0.05$   \\
Volatility of volatility  & $v = 0.1$     \\
Mean reversion speed      &     $\kappa = 2$\\
Long run mean of volatility  & $\theta = 0.1$    \\
Correlation   & $\rho = -0.5$   \\
Parabolic mesh ratio  & $\gamma = 0.5$ \\
Stochastic volatility drift parameter & $\alpha = 0$ \\
Stochastic volatility diffusion parameter & $\beta = 0.5$ \\
\bottomrule
\end{tabular}
\caption{Default input parameters for numerical experiments}
\label{defaultparams}
\end{center}
\end{table}

\begin{figure}
\centering
\includegraphics[width=0.75\textwidth]{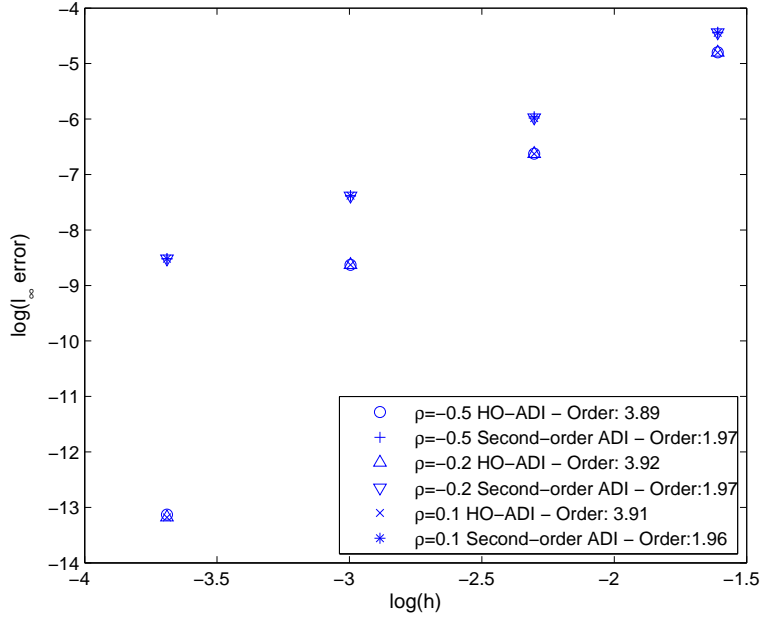}
\caption{$l_{\infty}$-error comparison of the high-order ADI scheme
  with standard second-order in space ADI scheme for various values of
  the correlation parameter $\rho$}
\label{fig:2}
\end{figure}

We compute the $l_2$-norm error $\varepsilon_2$ and the maximum norm
error $\varepsilon_\infty$ of the numerical solution with respect to a
numerical reference solution on a fine grid. We fix the parabolic mesh ratio $\gamma=\Delta_t/h^2$ to
a constant value which is natural for parabolic partial differential
equations as \eqref{PDEtransf}.
Then, asymptotically, we expect these
errors to converge as $\varepsilon = Ch^m$
for some $m$ and $C$ representing constants. This implies
$\ln(\varepsilon) = \ln(C) + m \ln(h) .$
Hence, the double-logarithmic plot $\varepsilon$ against $h$ should be
asymptotic to a straight line with slope $m$. This gives a method for
experimentally determining the order of the scheme. We expect to observe a numerical convergence rate of approximately order $\mathcal{O}(h^4)$ in space.
For comparison, we conduct additional
experiments using a standard, second-order ADI scheme based on
\eqref{HVscheme} combined with a 
second-order central difference discretisation in space.
Figure~\ref{fig:2} shows the double logarithmic plot of
$l_\infty$-error versus space step $h=\Delta_y=\Delta_x$.
We observe that the numerical convergence order agrees well with the
theoretical order of the schemes.
In all cases, the high-order ADI scheme outperforms the standard
second-order ADI scheme for a given mesh width $h$. Or in other words,
to realise a
chosen level of accuracy we could use a coarser grid 
for the high-order ADI scheme than the standard second-order scheme which translates into solving smaller linear systems and therefore is more computationally efficient. 

\subsection{Numerical stability analysis}

\noindent In this section we investigate whether there are any stability
restrictions on the choice of the time-step $\Delta_t$ for the
high-order ADI scheme. 
Unlike for standard second-order schemes, the algebraic complexity of
the numerical stability analysis of high-order compact schemes is
very high since the established stability 
notions imply formidable algebraic problems for high-order compact
schemes. As a result, there are only few stability results for
high-order compact schemes in the literature \cite{DuHe15,DuFo12,FournieRigal}. This is even more
pronounced in higher spatial dimensions, as most of the existing
studies with analytical stability results for high-order compact
schemes are limited to a one-dimensional setting. 

For diffusion
equations (without convection) with mixed derivative terms and
constant coefficients, a stability analysis of the ADI method
\eqref{HVscheme} with standard second-order discretisation in space
\cite{HouWel07} revealed it to be unconditionally stable. The analysis
in \cite{HouWel07}  is based on studying the stability for a simplified, linear test
equation which implies the assumption that 
all involved discretisation matrices are normal and commuting.
The discretisation matrices of high-order compact schemes generally do
not fulfil these assumptions and, hence, in the present case with
non-constant coefficients, the situation is much more involved. A
thorough stability analysis is therefore beyond the scope of the
present paper. Instead, to validate the stability of the scheme, we perform
additional numerical stability tests. We remark that in our numerical
experiments we observe stable behaviour throughout. 
We compute numerical solutions for varying values of the parabolic
mesh ratio $\gamma=\Delta_t/h^2$ and the mesh width $h$. Plotting the
associated $l_2$-norm errors in the plane should allow us to detect
stability restrictions depending on $\gamma$ or oscillations that
occur for high cell Reynolds numbers (large $h$). This approach for a
numerical stability study was also used in \cite{DuFo12,DuFoHe14}.

\begin{figure}
\centering
\includegraphics[scale = 0.45]{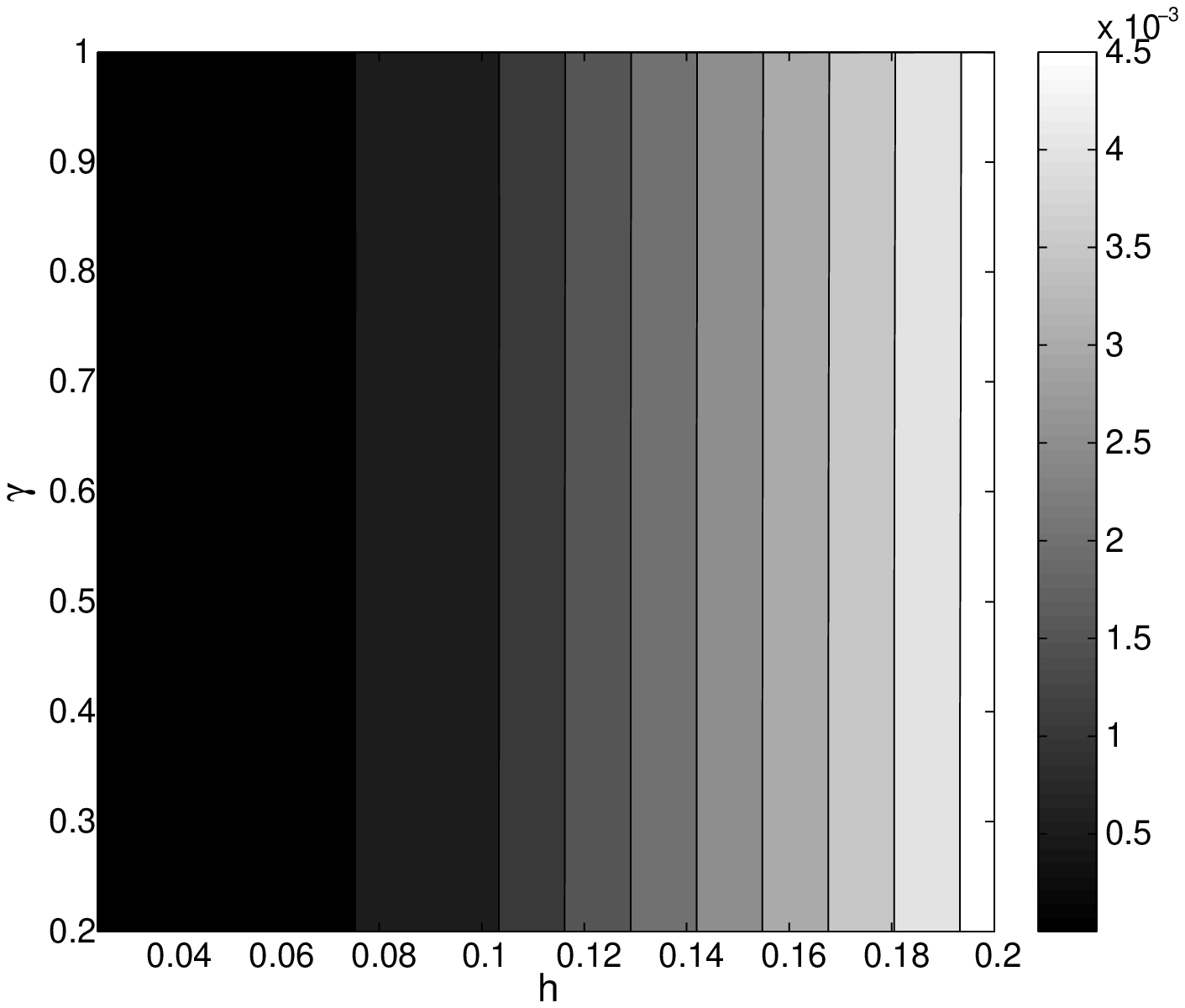}%
\includegraphics[scale = 0.45]{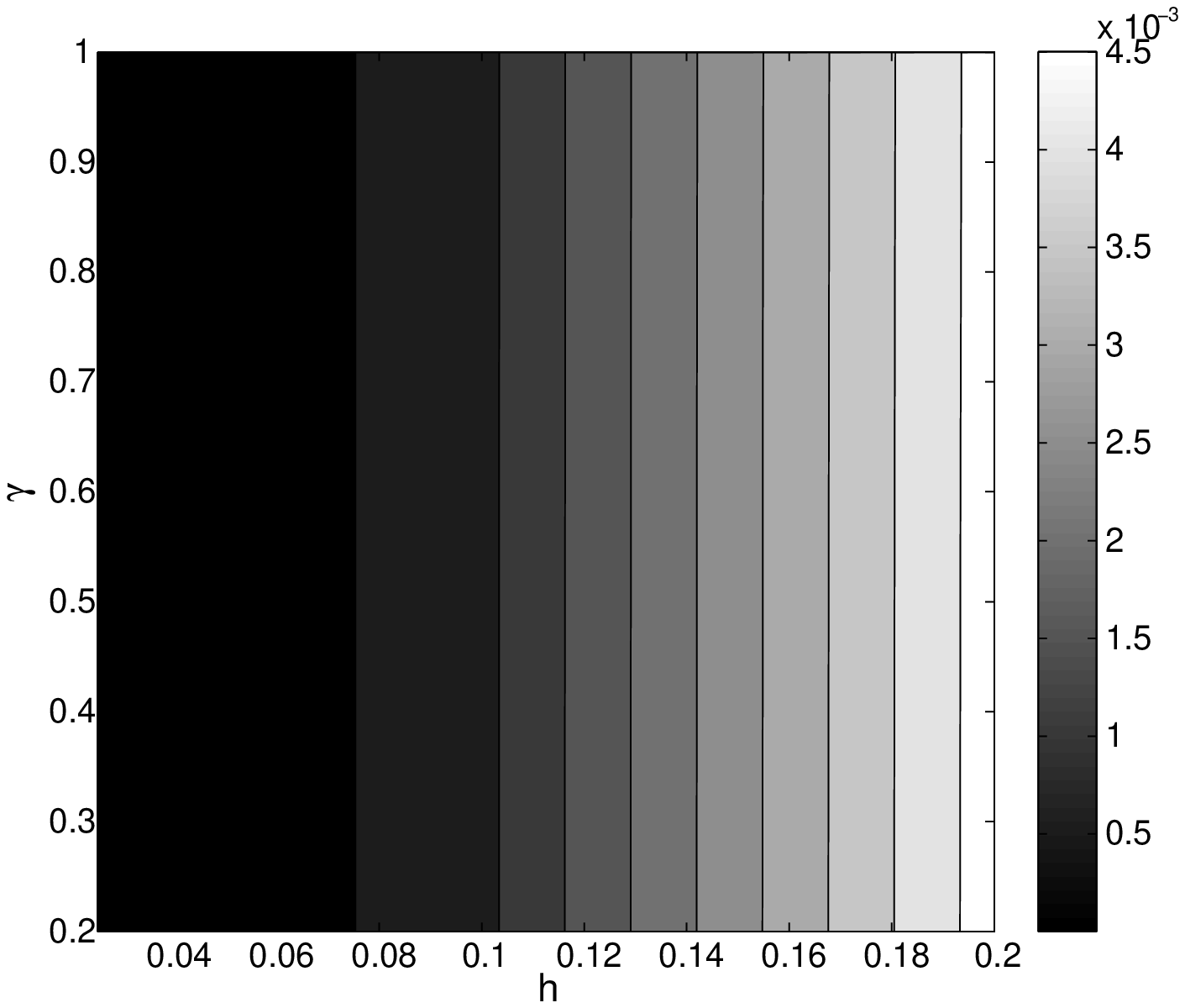}
\caption{Contour plot of the $l_2$-error for $\rho=0$ (left) and
  $\rho=-0.5$ (right) against parabolic mesh ratio $\gamma = {\Delta_{\tau}}/{h^2}$ and mesh width $h$}
\label{fig:3}
\end{figure}

We give results for the European Put option using the parameter from Table~\ref{defaultparams}. For our stability plots
we use $\gamma=k/10$ with $k=2,\ldots,10,$ and a descending sequence
of spatial grid points.
Figure~\ref{fig:3} shows the stability plots for the correlation
parameter $\rho = 0$ and  $\rho = -0.5$. 
We observe that
the influence of the parabolic mesh ratio $\gamma$ on the $l_2$-error
is only marginal and the relative error does not exceed $5 \times 10^{-3}$ as a
value for both stability plots. We can infer that there does not seem
to be a stability condition on $\gamma$ for either situation. For increasing
values of $h$, which also result in a higher cell Reynolds number, the
error grows gradually, and no oscillations in the numerical solutions
occur.  

These observations are confirmed by additional numerical convergence
tests for varying parabolic mesh ratio $\gamma$. The numerical
convergence orders reported in Table~\ref{tab:1} show that the
numerical convergence order for the high-order scheme, measured both
in the $l_2$-norm and $l_{\infty}$-norm is very close to four, and
does not depend on the parabolic mesh ratio $\gamma$.
\begin{table}
\centering
\begin{tabular}[l]{ l c c c c c}
\toprule
 $\gamma=\Delta_t/h^2$ & 0.2 & 0.4 & 0.6 & 0.8 & 1.0 \\
\midrule
 HO-ADI $l_2$-error & 3.8871 & 3.8870 & 3.8868 & 3.8866 & 3.8864 \\
 Standard ADI $l_2$-error & 2.4521 & 2.4519 & 2.4517 & 2.4514 & 2.4510 \\
 HO-ADI $l_{\infty}$-error & 3.8960 & 3.8961 & 3.8961 & 3.8962 & 3.8964\\ 
 Standard ADI $l_{\infty}$-error & 1.9744 & 1.9744 & 1.9744 & 1.9743 & 1.9742\\
\bottomrule 
\end{tabular}
\caption{Numerical convergence order in space for varying parabolic mesh ratio $\gamma=\Delta_t/h^2$}
\label{tab:1}
\end{table}

\section{Conclusion}
\label{sec:conc}

\noindent By combining fourth-order (compact and non-compact) finite difference
schemes in space with Hundsdorfer and Verwer's second-order ADI time-stepping scheme, we have constructed a new numerical method for solving option pricing problems for stochastic volatility models. Numerical experiments for approximating the price of a European Put option using Heston’s stochastic volatility model with generic parameters confirm the numerical convergence of the scheme in space and time while the results for a range of parabolic mesh ratios suggest unconditional stability.

 \section*{Acknowledgement}
\noindent BD acknowledges partial support by the Leverhulme Trust research project grant `Novel discretisations for higher-order nonlinear PDE' (RPG-2015-69).
JM has been supported in part by a studentship under the EPSRC Doctoral Training Partnership (DTP) scheme.


\end{document}